\documentclass[openacc]{rstransa}
\usepackage{amssymb,amsmath,upgreek,bm}
\usepackage[shortlabels]{enumitem}

\newtheorem{principle}{\bf Principle}[]
\newcommand{\beq}{\begin{equation}}
\newcommand{\eeq}{\end{equation}}
\newcommand{\beqa}{\begin{eqnarray}}
\newcommand{\eeqa}{\end{eqnarray}}
\newcommand{\beqan}{\begin{eqnarray*}}
\newcommand{\eeqan}{\end{eqnarray*}}

\newcommand{\ket}[1]{|{#1}\rangle}

\newcommand{\ip}[1]{\langle{#1}\rangle}

\newcommand{\eq}[1]{Eq.~(\ref{#1})}
\newcommand{\eqs}[1]{Eqs.~(\ref{#1})}
\newcommand{\eqr}[1]{(\ref{#1})}

\newcommand{\ii}{{\rm i}}
\newcommand{\dd}{{\rm d}}

\begin{document}

\title{The quantum theory of time, the block universe and human experience
}

\author{Joan A. Vaccaro}

\address{Centre for Quantum Dynamics, Griffith University, Nathan 4111 Australia}
	
\subject{quantum physics}


\corres{J.A. Vaccaro\\
	\email{J.A.Vaccaro@griffith.edu.au}}

\begin{abstract}
Advances in our understanding of the physical universe have impacted dramatically on how we view ourselves.
Right at the core of all modern thinking about the universe is the assumption that dynamics is an elemental feature that exists without question.
However, ongoing research into the quantum nature of time is challenging this view: my recently-introduced quantum theory of time suggests that dynamics may be a phenomenological consequence of a fundamental violation of time reversal symmetry.
I show here that there is consistency between the new theory and the block universe view.
I also discuss the new theory in relation to the human experience of existing in the present moment, able to reflect on the past and contemplate a future that is yet to happen.
\end{abstract}


\begin{fmtext}

Advances in science have had a profound impact on how we view ourselves.
In the 1500's the Copernican model of the solar system overturned the Ptolemaic model and with it the notion that human activity was a centrepiece of the cosmos.
Over the course of the next three centuries the development of classical mechanics led to the view of a clockwork universe that operated deterministically.
The extent of the determinism was extreme to the point considered by Laplace where a being (i.e. Laplace's demon) with sufficient resources could, in principle, predict the entire future of the universe from knowledge of its present state \cite{Weinert}.  The notion that we, as a part of the universe, can act freely was challenged as a result.
A new challenge emerged a century ago with Minkowski's spacetime \cite{Minkowski} as it provided support for viewing the universe as a 4-dimensional spacetime block that exists as one entity.
In this view, called the \emph{block universe} (or \emph{eternalism} in philosophical discussions), there is no basis for singling out a present time that separates the past from the future because all times coexist with equal status.
As such, it has profound implications for how we view our experience of existing in the present moment, able \linebreak
\end{fmtext}


\maketitle
\noindent
to reflect on the past and plan new actions in a variable future.
Indeed, if all times coexist with equal status, the future is as fixed as the past and so humans can only act in compliance with the future; this directly conflicts with the notion that humans can act freely.

However, the discovery of quantum mechanics in the last century has left puzzling implications in its wake, particularly in relation to the status of local realism, which asserts that effects are local in the sense of propagating no faster than the speed of light and physical properties are real irrespective of whether they are measured or not \cite{EPR,Bohr,Bell}.
The central issue is that quantum mechanics appears to allow for statistical effects that are instantaneous despite being distant from their causes, a situation that Einstein once referred to as ``spooky actions at a distance'' \cite{Born-Einstein}.
There is a lack of consensus as to how we should interpret this situation.
My own view is influenced by the debate on celestial motion from the 1500's which, to me, is analogous to the present situation.
The retrograde motion of the outer planets that arises in the Ptolemaic model can be considered as a modern-day analogue of the instantaneous spooky action associated with quantum mechanics in the following way.
The retrograde motion is not fundamental, in that it is not the actual motion of the planets, but rather it is how the motion of the planets appear from the \emph{perspective} of the Earth, which has a \emph{special status} of being the centerpiece in the Ptolemaic model. In the same manner, the instantaneous spooky action should be considered not as fundamental, but rather as a result of taking a particular \emph{perspective} where humans (or certain other systems) are given the \emph{special status} of being essentially independent of the rest of the universe.
The special statuses and corresponding perspectives are removed on adopting the Copernican model and Brans' fully causal model \cite{Brans,Hall} in the historic and modern-day cases, respectively.
Moreover, the retrograde motion is explained away as being apparent motion and the instantaneous spooky action is accounted for by a common cause \cite{Brans,Cavalcanti}, respectively.
Further details of this argument can be found in the \ref{Ap:my view}.
The point I would like to make here is this.
Although Laplacian determinism challenges the notion that humans can act freely, insisting humans are not shackled by Laplacian determinism does not actually clash with classical physics because the physical actions of any human can be mimicked by an appropriately programmed and sufficiently dexterous robot entirely in accord with classical physics.
But now, following the discovery of quantum mechanics, we find that insisting human actions are essentially independent of the rest of the universe has an apparent physical consequence in the appearance of instantaneous spooky action.
The fact that this consequence can be placed in analogy with the apparent retrograde motion associated with the Ptolemaic model undermines the argument for human independence.
As such, it poses a new challenge to the notion that humans can act freely.

The tendency for scientific advances to reinforce the case for a mechanistic model of ourselves shows no sign of abatement.
In the following I discuss how my recently-introduced quantum theory of time \cite{PRSA,BookCh} maintains this trend.
In particular, the new theory involves a set of states, each one of which represents a physical system such as a galaxy as existing, in the case of time reversal symmetry, at one specific time.
However, in the case of a violation of time reversal symmetry (T violation) of the kind observed in particle physics \cite{Angelopoulos,Lees}, the set of states represent the galaxy as existing at a sequence of increasing times.
Each state provides ontological support our subjective perception of existing in the present and the ordered sequence of states provides ontological support for our perception of progressing through time from the past to the future.
The theory may thus appear to be at odds with the block universe for which events at all times coexist as one entity.
However, I argue below that the new theory is, in fact, consistent with the block universe view.
I show that the ordered sequence of states provides a physical basis for some of our subjective notions about time to be incorporated into the block universe view.
It resolves, therefore, the open problem of how to square the human perspective of time within the block universe \cite{Price-flow of time}.
Yet despite this, it offers no escape from determinism and the mechanistic human.

The organisation of the remainder of this paper is as follows.
I rework the original development of the new theory \cite{PRSA} into a set of four basic principles in section \ref{sec:principles} and summarize the main results of their implementation in section \ref{sec:main results}.
Next, in section \ref{sec:block universe}, I interpret the new theory in terms of the block universe and then, in section \ref{sec:implications}, I discuss the implications for how we experience time.
I end with a conclusion.

\section{Principles of the quantum theory of time\label{sec:principles}}

The development of the new theory \cite{PRSA,BookCh} can be distilled into four basic principles, each of which originates from a specific observation.
The first observation is that the accepted convention in physics is that dynamics is assumed to be an elemental part of nature---as existing without question---and is incorporated into physical theories through conservation laws and equations of motion.
However, this convention treats time differently to space because, in essence, systems evolve and conservation laws apply over time and not space.
This asymmetric treatment is at odds with the structure of Minkowski's spacetime which expresses an equivalence between time and space \cite{Minkowski}.
If this time-space asymmetry was found to be phenomenological in origin, we might gain a better understanding of the nature of dynamics and time itself.
Accordingly, we begin with a representation of the quantum state of a system that treats time and space symmetrically by adopting the following:
\begin{principle}  \label{p:time-space symmetry}
     The physically-realisable states of a system have the same construction in both time and space. Any differences between space and time, such as dynamics and conservation laws, emerge phenomenologically.
\end{principle}
\noindent
The important point here is that it allows a physical system to be represented as being localised in both space and time.
The kinds of systems being considered could be any system describable in conventional quantum mechanics.
Allowing localisation in time rules out the possibility of incorporating equations of motion and conservations laws at a fundamental level of the theory.

Next, we note that the generators of translations in time and space are given by the Hamiltonian $\hat H$ and the momentum operators $\hat {\mathbf{P}}$, respectively.
They are distinguished by the discrete symmetries of charge conjugation (C), parity inversion (P) and time reversal (T) in the sense that only the temporal generator expresses violations of the symmetries.
In fact, the violations are only seen in the time evolution of various particle decays and thus in translations over time.
This suggests that the discrete symmetry violations may provide a phenomenological basis for the kind of time-space asymmetry needed for dynamics to emerge in accord with Principle \ref{p:time-space symmetry}.
In particular, if $\hat H$ represents the Hamiltonian for evolving into the future, then the Hamiltonian for the time reversed evolution into the past is $\hat{\mathcal{T}}^{-1}\hat H\hat{\mathcal{T}}$  where $\hat{\mathcal{T}}$ is Wigner's time reversal operator \cite{Wigner}.
For brevity, we use a special symbol for each version of the Hamiltonian, viz.
\begin{align}
       \hat{H}_F &=\hat{H}\ ,\qquad \hat{H}_B =\hat{\mathcal{T}}^{-1}\hat H\hat{\mathcal{T}}
\end{align}
where the subscripts ``F'' and ``B'' represent the ``forwards'' (or positive $t$) and ``backwards'' (negative $t$) directions of time evolution, respectively.
T violation is represented by
\begin{align}
        \hat{H}_F \ne \hat{H}_B\ .
\end{align}
This gives rise to a seemingly ambiguous situation where there are twice as many unitary time translation operators than needed.
The following Principle provides the necessary clarification of the role of each type of operator; it's justification can be found in Ref. \cite{PRSA}.
\begin{principle} \label{p:time evolution}
    Physical time evolution is represented by the operators\footnote{For convenience, we use units for which $\hbar=1$.} $\exp(-\ii \delta t\hat H_F)$ and $\hat{\mathcal{T}}^{-1}\!\exp(-\ii \delta t\hat H_F)\hat{\mathcal{T}}=\exp(\ii \delta t\hat H_B)$ for the forward (positive $t$) and backward (negative $t$) directions of time, respectively, where $\delta t>0$ represents an interval of time $t$. The operators $\exp(\ii \delta t\hat H_F)$ and $\exp(-\ii \delta t\hat H_B)$ represent the mathematical inverse operations of  ``unwinding'' or ``backtracking'' the evolution represented by $\exp(-\ii \delta t\hat H_F)$ and $\exp(\ii \delta t\hat H_B)$, respectively.
\end{principle}
\noindent
The situation for spatial translations involves no such subtlety.
The operator representing a translation by a distance $\delta r$ in the direction of unit vector $\mathbf{n}$ is given by  $\exp(-\ii \delta r\hat{\mathbf{P}}\cdot \mathbf{n})$;  the corresponding operator for a translation in the reverse direction is simply  $\hat{\mathcal{P}}^{-1}\!\exp(-\ii \delta r\hat{\mathbf{P}}\cdot \mathbf{n})\hat{\mathcal{P}}=\exp(\ii \delta r\hat{\mathbf{P}}\cdot \mathbf{n})$ where $\hat{\mathcal{P}}$ is the parity inversion operator.
Note that these two spatial translation operators are inverses of each other.

The third observation is about Feynman's sum-over-paths (or path integral) formalism \cite{Feynman}.
Although the formalism may be considered a mathematical tool for calculating probability amplitudes, Feynman's demonstration that it gives the origin of Hamilton's least action principle in classical mechanics suggests that the associated interference between multiple paths has \emph{ontological significance} even to the extent of underlying the large scale classical structure of the universe.
As such, sums over paths should feature at a fundamental level in quantum theory.
Paths over space are traversed by application of the spatial translation operators $\exp(-\ii \delta r\hat{\mathbf{P}}\cdot \mathbf{n})$ and $\exp(\ii \delta r\hat{\mathbf{P}}\cdot \mathbf{n})$ for $\mathbf{n}=\frac{1}{\sqrt{3}}(j,k,\ell)$ where the values of $j$, $k$, and $\ell$ are randomly chosen from the set $\{1,-1\}$.
Note that the spatial translation operators are mutually commuting.
Paths over time are traversed by application of the time evolution operators $\exp(-\ii \delta t\hat H_F)$ and $\exp(\ii \delta t\hat H_B)$ for $\delta t>0$ according to Principle \ref{p:time evolution}, and so paths with reversals in direction contain both types of time evolution operator.
More reversals implies more instances of each kind of evolution operator.
A particularly important case is where $\exp(-\ii \delta t\hat H_F)$ and $\exp(\ii \delta t\hat H_B)$ are non-commuting because then paths over time become quite distinct from paths over space.
This phenomenological asymmetry between time and space is the kind we seek, and so we adopt the following:
\begin{principle}  \label{p:qvp}
    Let $W$ represent a temporal or spatial axis and $w$ represent a coordinate value on it.
    Let $\ket{w_0}_{\rm w}$ be a ket that is extremely localised near $w=w_0$ to the extent that it is not representative of any physically-realisable state.\footnote{This makes use of the fact that a physical constraint limits the states that a system may occupy but it doesn't limit the basis states that may be used to give expanded representations of any physically-realisable state.
    In particular, the constraint of bounded momentum restricts spatial distributions to having a minimum width, but allows a position eigenstate basis expansion.}
    We express physically-realisable kets $\ket{\phi_N(w_0,\sigma)}_{\rm  w}$ in terms of  $\ket{w_0}_{\rm w}$ as follows.
    Imagine translating $\ket{w_0}_{\rm w}$ over all possible paths along $W$ where each path comprises $N$ steps, each step is a translation by the interval
    \begin{align}
            \label{eq:delta w definition}
            \delta w_N=\frac{\sigma}{\sqrt{N}}
    \end{align}
    in either $\pm w$ direction, and $\sigma$ is a width parameter.
    The translation of $\ket{w_0}_{\rm  w}$ over each path will result in an extremely-localised ket $\ket{w'}_{\rm  w}$ at a corresponding final coordinate value $w=w'$.
    The ket $\ket{\phi_N(w_0,\sigma)}_{\rm  w}$ is defined to be the superposition of all such translated kets according to
    \begin{align}
        \label{eq:QVP over W defn}
        \ket{\phi_N(w_0,\sigma)}_{\rm  w}
        &\equiv
             \frac{1}{2^N}\left(\hat{\mathcal{Q}}^{-1}e^{-\ii\hat{G}_+\delta w_N}\hat{\mathcal{Q}}+e^{-\ii\hat{G}_+\delta w_N}\right)^N\ket{w_0}_{\rm w}\ ,\\
        \label{eq:QVP over W}
        &=\frac{1}{2^N}\left(e^{\ii\hat{G}_-\delta w_N}+e^{-\ii\hat{G}_+\delta w_N}\right)^N\ket{w_0}_{\rm w}
    \end{align}
    where $\hat{\mathcal{Q}}$ is the discrete symmetry operator associated with $W$ (i.e. either the time reversal or the parity inversion operator), and $\hat{G}_\pm$ is the generator of translations along $W$ in the $\pm w$ directions, respectively.
    The right side of \eq{eq:QVP over W defn} is called a ``quantum virtual path (QVP)'' over $W$; it expresses the ket $\ket{\phi_N(w_0,\sigma)}_{\rm  w}$ in terms of a mathematical construction that is regarded in the theory as being fundamental.
\end{principle}
\noindent
QVPs extend Feynman's sum-over-paths formalism \cite{Feynman} in two different ways.
The first is that, whereas each of Feynman's paths are, in general, possible trajectories in space and time, here each path represents a sequence of \emph{virtual displacements} in either space or time.
Virtual displacements arise in analytical mechanics when discussing constraints on motion \cite{Goldstein}.
A QVP is the superposition of all possible paths of this kind (where the adjective ``quantum'' is used in the same sense as in the label ``quantum walks'' \cite{quantum-walk}).
The second is that, whereas Feynman used sums over paths as a means for calculating the transition amplitude for a process with respect to an initial and a final state, here a QVP is associated with the \emph{quantum state of a system} itself.

The final observation that we need to construct the new theory is that there are expected to be fundamental limits to the degree to which intervals in time and space can be probed.
For example, quantum gravity is expected to limit the resolution to the Plank time $\approx10^{-44}~$s and the Planck length $\approx 10^{-35}~$m \cite{Hossenfelder,Garay,DAriano}.
This implies that QVPs over the $W$ axis with step sizes $\delta w_N$ in \eq{eq:delta w definition} that are smaller than the corresponding fundamental limit $\delta w_{\rm min}$ should be accorded the same physical status in representing the state of the system.
Even though the expected limits are extremely small, they allow for an unbounded number of QVPs of equivalent physical status with $\delta w_N<\delta w_{\rm min}$.
To account for this, we adopt the following principle:
\begin{principle}  \label{p:resolution}
There is a lower limit to the resolution of intervals in space and time.  QVPs with step sizes smaller than this limit have an equal physical status in representing the state of the system.
The set of all QVPs with equal physical status is given by
\begin{align}  \label{eq:set Phi W}
        \bm{\Upphi}_{w_0,\sigma}=\{\ket{\phi_N(w_0,\sigma)}_{\rm w}: N\ge N_{\rm w}\}
\end{align}
where $N_{\rm w}$ is the smallest value of $N$ in \eq{eq:delta w definition} that gives $ \delta w_N<\delta w_{\rm min}$.
Each QVP in $\bm{\Upphi}_{w_0,\sigma}$ is associated with identical values of the physical parameters $\sigma$ and $w_0$ and is distinguished mathematically by a unique value of the physically-inaccessible index $N$.
The set $\bm{\Upphi}_{w_0,\sigma}$ itself is, therefore, a representation of the state of the system characterised by the parameters $w_0$ and $\sigma$.
\end{principle}
\noindent
The underlying concept here is closely related to the construction of the set of real numbers:
representing a state by the set $\bm{\Upphi}_{w_0,\sigma}$ of QVPs with equivalent physical status is analogous to a real number being represented by an equivalence class of Cauchy sequences of rational numbers \cite{Ackoglu}.

\section{Overview of the main results\label{sec:main results}}

We turn now to the main results that follow from implementing the four Principles; further details can be found in the original work \cite{PRSA}.
Consider first the situation where there is only one generator of translations over the generic axis $W$, i.e. where $\hat{G}_+=\hat{G}_-=\hat{G}$ in \eq{eq:QVP over W}.
Making use of the result $\exp(-A^2/2)=\lim_{N\to\infty}\cos^N(A/\sqrt{N})$ shows that every QVP in $\bm{\Upphi}_{w_0,\sigma}$ is physically indistinguishable from the limit ket $\ket{\phi_{\rm lim}(w_0,\sigma)}_{\rm w}=\exp(-\hat{G}^2\sigma^2/2)\ket{w_0}_{\rm w}$.

The one-generator situation occurs for the 1-dimensional spatial case where $\hat{G}$ corresponds to the momentum operator $\hat{P}_x$ in the $x$ direction, say.
In that case, we replace the generic label $\phi$ with $\psi$ and note that $\ket{w_0}_{\rm w}$ corresponds to the position eigenstate $\ket{x_0}_{\rm x}$, the set $\bm{\Upphi}_{w_0,\sigma}$ corresponds to
\begin{align}
        \bm{\Uppsi}_{x_0,\sigma}=\{\ket{\psi_N(x_0,\sigma)}_{\rm x}: N\ge N_{\rm x}\}
\end{align}
and $N_{\rm x}$ is the smallest value of $N$ that gives $ \delta x_N=\sigma/\sqrt{N}<\delta x_{\rm min}$ where $\delta x_{\rm min}$ is the fundamental spatial resolution limit.
Each QVP in $\bm{\Uppsi}_{x_0,\sigma}$ is found to be physically indistinguishable from the Gaussian-distributed limit ket
\begin{align}
        \ket{\psi_{\rm lim}(x_0,\sigma)}_{\rm x}=\int \dd x \exp[-(x-x_0)^2/(2\sigma^2)]\ket{x}_{\rm x}\ .
\end{align}
The set $\bm{\Uppsi}_{x_0,\sigma}$ is redundant in the sense that it contains infinitely-many QVPs that are equally able to represent the state of an object that is located near $x=x_0$ within a region of order $\sigma$.
The set $\bm{\Uppsi}_{x_0,\sigma}$ itself is, therefore, a representation of the 1-dimensional spatial state of the object in accord with Principle \ref{p:resolution}.

The one-generator situation also occurs for QVPs over time in the case of T symmetry where $\hat{G}$ corresponds to $\hat{H}_F=\hat{H}_B=\hat{H}$, and so an equivalent situation results.
Using $\Upsilon$ instead of $\phi$ as the label gives the set corresponding to \eq{eq:set Phi W} as
\begin{align}  \label{eq:set Upsilon T symmetry}
        \bm{\Upupsilon}_{t_0,\sigma}=\{\ket{\Upsilon_N(t_0,\sigma)}_{\rm t}: N\ge N_{\rm t}\}
\end{align}
where $N_{\rm t}$ is the smallest value of $N$ that gives $ \delta t_N=\sigma/\sqrt{N}<\delta t_{\rm min}$ where $\delta t_{\rm min}$ is the fundamental temporal resolution limit.
Just as in the spatial case, each ket in $\bm{\Upupsilon}_{t_0,\sigma}$ is found to be physically indistinguishable from the Gaussian-distributed limit ket
\begin{align}
        \ket{\Upsilon_{\rm lim}(t_0,\sigma)}_{\rm t}=\int \dd t \exp[-(t-t_0)^2/(2\sigma^2)]\ket{t}_{\rm t}\ .
\end{align}
Two comments are in order here.
First, the ket $\ket{t}_{\rm t}$ is related to a fixed ket $\ket{t_0}_{\rm t}$ by time translation, i.e. $\ket{t}_{\rm t}=\exp[-\ii \hat{H}(t-t_0)]\ket{t_0}_{\rm t}$.
Second, the fixed ket $\ket{t_0}_{\rm t}$ is assumed to be well-defined in time in the sense that translating it in time by a small amount yields, essentially, an orthogonal state, viz. the value of ${}_{\rm t}\kern-0.1em\ip{t|t_0}_{\rm t}$ is negligible except near $t=t_0$; this implies that $\ket{t_0}_{\rm t}$ has a correspondingly very broad energy distribution.

The set $\bm{\Upupsilon}_{t_0,\sigma}$ represents the galaxy as existing temporally only for a duration of the order of $\sigma$ around the time $t_0$ in the same way that $\bm{\Uppsi}_{x_0,\sigma}$ represents it as existing spatially in a region of order $\sigma$ near the spatial point $x_0$.
Clearly, these states have the same construction in time as in space in accord with Principle \ref{p:time-space symmetry}.
As a consequence, the state $\bm{\Upupsilon}_{t_0,\sigma}$
represents the violation of the conservation of mass and energy.
There is also no equation of motion inherent in the theory to represent time evolution of the galaxy over extended periods of time.
The study of states that are localised in time like this has a history dating back to Stueckelberg in the 1940's and Feynman, Nambu and Schwinger in the 1950's \cite{Fanchi}.
However, in contrast to that here, these previous studies incorporated dynamical equations as an integral part of their approaches and inherently assumed time evolution.

Next, consider the situation where there are two distinct generators, i.e. $\hat{G}_+\ne \hat{G}_-$.
This occurs for QVPs over time when T violation is present, i.e. $\hat{H}_F\ne\hat{H}_B$.
Let $\hat{G}_+$ and $\hat{G}_-$ in \eq{eq:QVP over W} correspond to $\hat{H}_F$ and $\hat{H}_B$, respectively.
We assume the commutator of the two versions of the Hamiltonians to be of the simple form
\begin{align}
   \label{eq:commutator}
   [\hat H_{\rm B},\hat H_{\rm F}]=\ii\lambda\ .
\end{align}
The resulting QVPs depend on the physical parameter $\lambda$ in addition to $t_0$ and $\sigma$, and so we need to include it in expressions accordingly.
Thus, in place of \eq{eq:set Upsilon T symmetry}, we now have
\begin{align} \label{eq:set qvt time T violation}
        \bm{\Upupsilon}_{t_0,\sigma,\lambda}=\{\ket{\Upsilon_N(t_0,\sigma,\lambda)}_{\rm t}: N\ge N_{\rm t}\}\ .
\end{align}
The non-commutativity of the time evolution operators  $\exp(-\ii \delta t\hat H_F)$ and $\exp(\ii \delta t\hat H_B)$ leads to destructive interference between the virtual paths to the extent that the QVPs in $\bm{\Upupsilon}_{t_0,\sigma,\lambda}$ are given approximately, on coarse graining over time, by \cite{PRSA}
\begin{align}\label{eq:Upsilon with two peaks}
     \ket{\Upsilon_N(t_0,\sigma,\lambda)}_{\rm t}\propto
        \exp(\ii\hat H^{\rm phen}_{\rm B}t_{\rm c})\ket{t_0}_{\rm t} +
             \exp(-\ii\hat H^{\rm phen}_{\rm F}t_{\rm c})\ket{t_0}_{\rm t}
\end{align}
where
\begin{align}
       \hat{H}^{\rm phen}_{\rm F}  &=\hat H_{\rm F}a_+ - \hat H_{\rm B}a_-=\hat{\mathcal{T}}\hat{H}^{\rm phen}_{\rm B}\hat{\mathcal{T}}^{-1}\ ,\\
       a_\pm &= \frac{\theta}{4\pi}\pm \frac{1}{2}\ ,\\
       \theta &= \sigma^2\lambda\ ,\label{eq:theta}\\
       t_{\rm c} &= \frac{2\pi}{\sigma\lambda}\sqrt{N} \label{eq:clock time}
\end{align}
for $2\pi <\theta < 4\pi$.
Each term on the right of \eq{eq:Upsilon with two peaks} represents a modified QVP that has different contributions of time translation by $\hat{H}_{\rm F}$ and $\hat{H}_{\rm B}$ in accord with Principle \ref{p:time evolution}.
The operators $\hat{H}^{\rm phen}_{\rm F}$ and $\hat{H}^{\rm phen}_{\rm B}$ are the corresponding weighted averages of $\hat{H}_{\rm F}$ and $\hat{H}_{\rm B}$ and represent the phenomenological Hamiltonians that would be observed within the galaxy.
They can be thought of as the effective generators of the net time translation in each term on the right side of \eq{eq:Upsilon with two peaks}.
The parameter $t_{\rm c}$ represents the corresponding net translation in time that would be observed within the galaxy using conventional clocks.\footnote{The parameter $t_{\rm c}$ should be written as a function of $N$ in line with \eq{eq:clock time}, viz. $t_{\rm c}(N)$, however as this makes the notation relatively cumbersome, we leave the dependance on $N$ implicit.}
The particular value $\theta=2.23\pi$ corresponds, approximately, to each term on the right of \eq{eq:Upsilon with two peaks} being associated with the minimum uncertainty in time that is allowed by the minimum possible uncertainty in energy \cite{PRSA supp mat}.

The ket $\ket{\Upsilon_N(t_0,\sigma,\lambda)}_{\rm t}$ in \eq{eq:Upsilon with two peaks} represents the galaxy existing in a superposition of the two times $t_0\pm t_{\rm c}$ which reflects the symmetry inherent in the definition of QVPs in \eq{eq:QVP over W defn}.
Selecting increasing values of $N$ gives a sequence of kets $\ket{\Upsilon_N(t_0,\sigma,\lambda)}_{\rm t}$ in $\bm{\Upupsilon}_{t_0,\sigma,\lambda}$ associated, according to \eq{eq:clock time}, with increasing values of $t_{\rm c}$.
One take of this situation, based on the Feynman-Stueckelberg interpretation of antiparticles as particles travelling backward in time, is that the galaxy is matter-dominator at one of the times, say $t_0+t_{\rm c}$, and antimatter-dominated at the other time, $t_0-t_{\rm c}$.
Bi-directional time evolution like this has been explored previously by Carroll, Barbour and coworkers \cite{Carroll,Barbour}.
To see what it means here, consider a mechanical clock based on the escapement mechanism with hands that rotate clockwise with respect to increasing values of $t_{\rm c}$ and constructed from matter in the matter version of the galaxy at time $t_0+t_{\rm c}$; a little thought will reveal that a corresponding clock device in the antimatter version of the galaxy at time $t_0-t_{\rm c}$, that is equivalent in every respect (including the shape of the teeth on the escape wheel) except being constructed from antimatter, will have hands that also rotate \emph{clockwise} with increasing values of $t_{\rm c}$.
This implies that if both clocks are synchronised at one particular value of $t_{\rm c}$, they will remain in synchrony for larger values of $t_{\rm c}$.
As the value of the parameter $t_0$ that labels the ket $\ket{t_0}$ is arbitrary, we set
\begin{align}
        t_0=0
\end{align}
for convenience.
We assume all clocks, irrespective where they are located, are synchronised to display $t_{\rm c}$, which hereafter we refer to as the \emph{clock time}.
Also, for brevity, the location of the galaxy will only be quoted explicitly for positive values of time, leaving the corresponding situation for negative values implicit, unless clarity warrants otherwise.
The values of $t_{\rm c}$ that correspond to all the kets in $\bm{\Upupsilon}_{t_0,\sigma,\lambda}$ is found from \eqs{eq:Upsilon with two peaks} and \eqr{eq:clock time} to cover an unbounded range starting from the minimum value of $t_{\rm c, min}=2\pi\sqrt{N_{\rm t}}/(\sigma\lambda)$.
Interpreting $\bm{\Upupsilon}_{t_0,\sigma,\lambda}$ as a state in accord with Principle \ref{p:resolution} implies, therefore, that the galaxy exists for all times except in the interval $(-t_{\rm c, min},t_{\rm c, min})$.
Outside this interval the galaxy is represented as existing essentially continuously and, in this sense, conservation of mass arises as a property of $\bm{\Upupsilon}_{t_0,\sigma,\lambda}$.
The dynamics of the galaxy follows the Schr\"{o}dinger equation in the sense that
\begin{align}
        \frac{\dd}{\dd (-t_{\rm c})}\exp(\ii\hat H^{\rm phen}_{\rm B}t_{\rm c})\ket{t_0}_{\rm t} &=-\ii\hat H^{\rm phen}_{\rm B}\exp(\ii\hat H^{\rm phen}_{\rm B}t_{\rm c})\ket{t_0}_{\rm t}\ ,\\
        \frac{\dd}{\dd t_{\rm c}}\exp(-\ii\hat H^{\rm phen}_{\rm F}t_{\rm c})\ket{t_0}_{\rm t} &=-\ii\hat H^{\rm phen}_{\rm F}\exp(-\ii\hat H^{\rm phen}_{\rm F}t_{\rm c})\ket{t_0}_{\rm t}\ ,
\end{align}
and so the first and second terms on the right side of \eq{eq:Upsilon with two peaks} represent time evolution in the backward and forward directions of time, respectively.
Moreover, the set $\bm{\Upupsilon}_{t_0,\sigma,\lambda}$ is ordered because, in line with Principle \ref{p:time evolution}, $\ket{A}\equiv\exp(-\ii\hat H^{\rm phen}_{\rm F}t_{\rm c})\ket{t_0}_{\rm t}$ is related to $\ket{B}\equiv\exp(-\ii\hat H^{\rm phen}_{\rm F}t'_{\rm c})\ket{t_0}_{\rm t}$ for $t'_{\rm c}>t_{\rm c}$ by a net time \emph{evolution} according to\footnote{The net effect of time evolution for a duration $a_+\Delta t_{\rm c}$ in the forward direction combined with time evolution for a duration of $a_-\Delta t_{\rm c}$ in the backward direction is a net evolution of $a_+\Delta t_{\rm c}-a_-\Delta t_{\rm c}=\Delta t_{\rm c}$ in the forward direction as $a_+>a_-$.}
\begin{align} \label{eq:net time evolution}
        \ket{B} \propto \exp(-\ii\hat H^{\rm phen}_{\rm F}\delta t_{\rm c})\ket{A}
        =\exp(-\ii\hat H_{\rm F}a_+\delta t_{\rm c} + \ii\hat H_{\rm B}a_-\delta t_{\rm c})\ket{A}
\end{align}
for $\delta t_{\rm c}=t'_{\rm c}-t_{\rm c}$, whereas the relation
\begin{align}
        \ket{A} \propto \exp(\ii\hat H^{\rm phen}_{\rm F}\delta t_{\rm c})\ket{B}
\end{align}
gives the \emph{rewinding} of \eq{eq:net time evolution}.
This justifies taking the ket $\ket{\Upsilon_{N+n}(t_0,\sigma,\lambda)}_{\rm t}$ to be correspondingly \emph{more evolved} in time compared to $\ket{\Upsilon_N(t_0,\sigma,\lambda)}_{\rm t}$ for $n>0$.
All of these results taken together show that the new theory offers a framework that gives the \emph{origin of dynamics}.

\section{Consistency with the block universe\label{sec:block universe}}

We are now in a position where we can interpret the new theory in terms of the block universe.
For this, we restrict our attention to T violation and the temporal case.
An individual ket such as $\ket{\Upsilon_N(t_0,\sigma,\lambda)}_{\rm t}\in \bm{\Upupsilon}_{t_0,\sigma,\lambda}$ describes the galaxy when its internal clock time is $t_{\rm c}$ according to \eq{eq:clock time}, and so it could be regarded as a \emph{conditional state} of the galaxy \emph{for a given value of} $t_{\rm c}$.
The full or unconditional state of the galaxy, according to Principle \ref{p:resolution}, is the whole set $\bm{\Upupsilon}_{t_0,\sigma,\lambda}$.
In particular, as all elements in $\bm{\Upupsilon}_{t_0,\sigma,\lambda}$ have equal physical status, there is no basis for choosing \emph{one} element as representing the ``present'' state of the galaxy such that it divides the remaining elements into those representing ``past'' or ``future'' times.
Moreover, $\bm{\Upupsilon}_{t_0,\sigma,\lambda}$ represents the galaxy evolving according to the phenomenological Hamiltonian associated with the corresponding direction of time as described by \eq{eq:Upsilon with two peaks}.
As such, the dynamics of the galaxy is fixed for all times apart for the small interval $(-t_{\rm c, min},t_{\rm c, min})$.
In this respect the new theory is consistent with the principles of the block universe view.

The new theory, however, includes an additional feature.
The direction of time in the conventional block universe is the one associated with increasing entropy.
Conventionally, only time symmetric laws are considered, and so any time asymmetry in entropy is necessarily associated with asymmetric temporal boundary conditions, i.e. with the universe having low entropy at one extreme time and high entropy at another \cite{PriceBook}.
In contrast, the set $\bm{\Upupsilon}_{t_0,\sigma,\lambda}$ has an inherent temporal order such that an element associated with larger values of $N$ is more evolved in time, as discussed above.
This implies that if we pick one element of $\bm{\Upupsilon}_{t_0,\sigma,\lambda}$, say $\ket{\Upsilon_N(t_0,\sigma,\lambda)}_{\rm t}$, it would contain evidence of having evolved from $\ket{\Upsilon_{N'}(t_0,\sigma,\lambda)}_{\rm t}$ for $N'<N$ but no evidence of having evolved from $\ket{\Upsilon_{N''}(t_0,\sigma,\lambda)}_{\rm t}$ for $N<N''$.
For example, if $\ket{\Upsilon_{N'}(t_0,\sigma,\lambda)}_{\rm t}$ represented the galaxy containing a gas cloud that consisted of atoms in excited energy states that will spontaneously emit into a surrounding low-temperature electromagnetic background then, without any energy source to maintain the excited atomic states, the gas would be represented by $\ket{\Upsilon_{N}(t_0,\sigma,\lambda)}_{\rm t}$ and $\ket{\Upsilon_{N''}(t_0,\sigma,\lambda)}_{\rm t}$ as being progressively less excited, respectively.
Putting this altogether, the full state $\bm{\Upupsilon}_{t_0,\sigma,\lambda}$ is a sequence of conditional states $\ket{\Upsilon_{N}(t_0,\sigma,\lambda)}_{\rm t}$ in order of their corresponding clock times $t_{\rm c}$ where, in general, each conditional state contains physical evidence of the prior time evolution represented by the conditional states $\ket{\Upsilon_{N'}(t_0,\sigma,\lambda)}_{\rm t}$ for $t'_{\rm c}< t_{\rm c}$ where $t'_{\rm c}=\frac{2\pi}{\sigma\lambda}\sqrt{N'}$ from \eq{eq:clock time}.

The direction of time here is not set by the entropic nature of the galaxy, but rather it is due to the degree of time evolution in accord with Principle \ref{p:time evolution} and this, in turn, determines the direction in which entropy increases.
As a result, the new theory allows the block universe view to be extended, accordingly, to one containing an inherent direction of time.

\section{Implications for humans\label{sec:implications}}

The new theory offers some support for our subjective notions about time.
In particular, although not evident from the coarse-grained approximation given in \eq{eq:Upsilon with two peaks}, the conditional state $\ket{\Upsilon_N(t_0,\sigma,\lambda)}_{\rm t}$ represents the galaxy in coordinate time $t$ as being spread over a small interval around $t=\pm t_{\rm c}$.
In fact, a more accurate expression for $\ket{\Upsilon_N(t_0,\sigma,\lambda)}_{\rm t}$ is a superposition of two Gaussian-distributed kets centred on the times $t=\pm t_{\rm c}$ and each one with a variance in time of
\begin{align}   \label{eq:spread in time of galaxy}
       (\Delta t_{\rm c})^2\approx 2/|\lambda\tan(\theta/4)|\ ,
\end{align}
where $\lambda$ and $\theta$ are given by \eqs{eq:commutator} and \eqr{eq:theta}, respectively \cite{PRSA}.
The fact that the galaxy is represented by $\ket{\Upsilon_N(t_0,\sigma,\lambda)}_{\rm t}$ as localised in time is a sufficient objective basis to support the subjective perception of existing at a ``present'' moment at time $t_{\rm c}$.
Moreover, in the previous section we found that each conditional state $\ket{\Upsilon_N(t_0,\sigma,\lambda)}_{\rm t}$ contains, in general, physical evidence of time evolution prior to $t_{\rm c}$, and this provides an objective basis for contemplative human observers within the galaxy to perceive recalling the ``past''.
In other words, whenever the galaxy exists as represented by $\ket{\Upsilon_N(t_0,\sigma,\lambda)}_{\rm t}$, contemplative human observers could perceive the present moment and recall the past.
As the conditional state $\ket{\Upsilon_N(t_0,\sigma,\lambda)}_{\rm t}$ represents the galaxy existing only for a small duration around $t_{\rm c}$, the subjective perception of the present moment and recalling the past is likewise confined to the same region of time.
However, as all elements of $\bm{\Upupsilon}_{t_0,\sigma,\lambda}$ have equal physical status, if one element has this representation, then they all do equally!
The only option, then, is for the set $\bm{\Upupsilon}_{t_0,\sigma,\lambda}$ to be an objective basis for humans to perceive the present moment and recall the past at \emph{all} associated times $t_{\rm c}$.\footnote{This does not conflict with the statement in section \ref{sec:block universe} that there is no basis in the new theory for dividing the elements of $\bm{\Upupsilon}_{t_0,\sigma,\lambda}$ into those representing past times, a present time, and future times.
We are associating here \emph{all} elements with the present, and each element has evidence of prior evolution.}

There is a further implication for us in the distribution of the conditioned states $\ket{\Upsilon_N(t_0,\sigma,\lambda)}_{\rm t}$ over time.
The increase in $t_{\rm c}$ for successive values of $N$ is given by $\delta t_{\rm c}^{\rm succ} \approx \pi/(\sigma\lambda\sqrt{N})$ which is less than $\frac{1}{2}\delta t_{\rm min}$ \cite{PRSA}, where $\delta t_{\rm min}$ is the temporal resolution limit, and so the set of values of the clock time $t_{\rm c}$ are physically indistinguishable from a continuous range.
The fact that the spread in time $\Delta t_{\rm c}$ given by \eq{eq:spread in time of galaxy} is independent of $N$ whereas $\delta t_{\rm c}^{\rm succ} \approx \pi/(\sigma\lambda\sqrt{N})$ decreases as $N$ increases implies that the conditional states $\ket{\Upsilon_N(t_0,\sigma,\lambda)}_{\rm t}$ for sufficiently-large, successive values of $N$ will overlap in time.
A given clock time $t_{\rm c}$ will, therefore, be associated with the ordered collection of conditional states such as $\{\ket{\Upsilon_{N'}(t_0,\sigma,\lambda)}_{\rm t}: |t_{\rm c}-t'_{\rm c}|\le \Delta t_{\rm c}\}$ where $t'_{\rm c}=\frac{2\pi}{\sigma\lambda}\sqrt{N'}$.
In other words, any particular clock time value $t_{\rm c}$ is associated with an ordered collection of conditional states.
To make a comparison with human experience, note that the time scale associated with conscious awareness is of the order of a second \cite{conscious thought}, which is many orders of magnitude greater than the intervals such as $\Delta t_{\rm c}$ considered here.
The subjective awareness of time is associated, therefore, with a set of many conditional states $\ket{\Upsilon_N(t_0,\sigma,\lambda)}_{\rm t}$ that overlap in time and are ordered in terms of increasing clock time.
These time-ordered conditional states of the galaxy carry a description of time-ordered neurological effects in the humans within it and this provides an objective basis for the subjective perception of a ``flow'' or ``passage'' of time.

Finally, there is also a severe constraint for us as well.
The new theory is deterministic in the sense that it obeys Hamiltonian dynamics and so it does not allow us to act freely nor to have an ability to determine the future, despite our strong subjective perceptions to the contrary.
In short, it offers us no escape from determinism.

\section{Conclusion}

Quantum mechanics has spawned the development of a myriad of advanced technologies that drive the modern western way of life.
However, the foundations of quantum mechanics offer something more, a reason to question how we view ourselves within in an objective landscape.
Indeed, the test of local realism formulated by Bell \cite{Bell} appears to be a new challenge to the notion that humans can act freely.

In this paper we explored the implications of the quantum nature of time for humans.
The structure of the recently-introduced quantum theory of time \cite{PRSA} was distilled into four basic principles and the main results that arise from their implication was also briefly reviewed.
An important feature of the new theory is that dynamics and time evolution are not assumed to be part of its structure, rather they emerge phenomenologically due to T violation.
The state of a physical system over time is given by a collection of conditional states, where each conditional state represents the system at a specific value of clock time.

We found the new theory is consistent with the block universe in that the conditional states are fixed by unitary evolution and that no \emph{one} conditional state can be singled out as representing a present that divides other conditional states into sets representing the past and the future.
However, the new theory has additional features.
One is that each conditional state is localised in time around a mean clock time value and so it can represent a human having the subjective perception of the present moment.
Another is that the collection of conditional states has a natural ordering in terms of the degree of time evolution, and as a result, each conditional state contains, in general, evidence of past evolution.
This supports the subjective perception of being able to recall the past and contemplate the future.
It also provides an objective basis for the subjective perception of a flow or passage of time.
As such it offers an alternative to Price's view that the subjective passage of time does not have an objective basis \cite{Price-flow of time}.
In short, the collection of conditional states can represent a human at every clock time as having the subjective perception of the present moment with memory of the past and being aware of the passage of time.
These results essentially extend the block universe view.
Finally, the natural ordering due to T violation suggests a connection with causality \cite{Pegg,Jeffers} but this is yet to be explored.

\newcounter{appendix}
\refstepcounter{appendix}
\renewcommand\thesection{Appendix}
\section{Humans and local realism\label{Ap:my view}}

Here I give further details of the tension between human independence and local realism by describing the full implications of a violation of Bell's theorem in subsection \ref{App:violation} and then outlining a logical fallacy in an argument for human independence in subsection \ref{App:independence}.

\subsection{Implications of a violation of Bell's theorem\label{App:violation}}

Central to the issue of the instantaneous spooky action mentioned in the main text is Bell's theorem \cite{Bell,CHSH}.
The proof of the theorem hinges on the partitioning the universe into two distinct parts and assuming each part has a distinct physical description.
One part, which I will label as ``LR'' (for local realistic), contains the system of primary interest, such as a pair of correlated spins at widely separated locations, and is \emph{assumed} to have a local realistic description.
The other part, which I will label as ``MS'' (for measurement settings), contains systems whose properties determine the setting of the measurement that is to be performed on the LR part at each location, and each measurement setting is \emph{assumed} to be independent of the LR part (i.e. it is assumed the properties of the two parts are not determined by a common cause \cite{Cavalcanti,Larsson}); this assumption is often referred to as measurement independence.
The conventional interpretation of an experimental demonstration of a violation of Bell's theorem is that nature is inconsistent with local realism \cite{Shalm,Hensen,Giustina,Handsteiner,Rosenfeld}.
However, this is not the whole story \cite{Brans,Larsson}: a violation of the theorem implies the logical complement of the assumptions it hinges on, namely that the LR-MS partitioning is not respected by nature, and so either the MS part does not give settings that are independent of the LR part, or the LR part does not have a local realistic description.

Experiments are typically designed with the independence of the MS part in mind and employ various forms of quantum random number generators to determine the measurement settings \cite{Shalm,Hensen,Giustina,Handsteiner,Rosenfeld}.
Nevertheless, the independence cannot be guaranteed, even using light from distant stars \cite{Handsteiner}, because the possibility of a common cause cannot be ruled out \cite{Larsson}.
Here we explore the specific case, as imagined by Bell and others \cite{Bell,WisemanNat,Shimony,Zeilinger,Gisin}, where the measurement settings are determined by experimentalists in the MS part.
In this case the implications of a violation is either
\begin{enumerate}[(i),topsep=0pt]
\item \label{option-not indep} the experimentalists' actions (in the MS part) are not independent of the LR part, or
  \item \label{option-indep of NLR universe} the LR part does not obey local realism.
\end{enumerate}
We have no physical grounds to favour either \ref{option-not indep} or \ref{option-indep of NLR universe} and so it is up to us to choose which possibility we prefer.
However, if we insist on the view that humans can act independently of the universe that surrounds them then the only compatible possibility is \ref{option-indep of NLR universe}.
This is essentially the interpretation of a violation that is advocated by a number of leading figures in this research area \cite{Bell,WisemanNat,Shimony,Zeilinger,Gisin} and so I will refer to it as the \emph{leading interpretation}.
Possibility \ref{option-not indep} has been referred to as ``superdeterminism'' \cite{Bell,WisemanNat,Shimony,Zeilinger,Gisin}, however as ``fully causal'' \cite{Brans} is a more fitting description, I shall refer to \ref{option-not indep} as the \emph{fully causal interpretation}.

The question of which interpretation is preferable has a precedent in the debate about celestial motion in the 1500's as follows.
The Ptolemaic model of the 1500's is analogous to the leading interpretation of a Bell theorem violation now: both give humans a preferred status (i.e. as a centrepiece in the cosmos in the Ptolemaic model and independent of the universe that surrounds them in the leading interpretation) and both have unexplained phenomena as a consequence (i.e. apparent retrograde motion of the outer planets in the Ptolemaic model and non-local-realistic behaviour as typified by Einstein's ``spooky actions at a distance'' remark \cite{Born-Einstein} in the leading interpretation).
Alternatively, the Copernican model in the 1500's is analogous to the fully causal interpretation: both eliminate the preferred status of humans and both account for the unexplained phenomena in a simple way.

\subsection{Argument that humans are necessarily independent\label{App:independence}}

When faced with deciding the fundamental properties of the universe, as in the case of assessing the implications of a violation of Bell's theorem, it is important to understand the true merits of any argument that might be put forward.
An argument that has been advocated by leading physicists is that humans are necessarily independent of the universe that surrounds them because the practice of science requires the independence of the experimenter from the subject of study \cite{Shimony,Zeilinger,Larsson}.
For example, Shimony \emph{et al.} \cite{Shimony} state that unless the experimenter and subject are independent we would need to abandon ``...the whole enterprise of discovering the laws of nature by experimentation'' and Zeilinger \cite{Zeilinger} claims that if the experimenter and subject were not independent ``...such a position would completely pull the rug out from underneath science.''
However, this argument contains a logical fallacy called an appeal to consequences.
Specifically, arguing for experimenter-subject independence on the basis that the alternative has undesirable consequences does not prove that experimenters are independent of their subjects.
Rather, the alternative may well be true, in which case we would need to deal with the consequences.

For example, imagine the experimenters eliminate all sources experimenter-subject dependence possible, and they perform experiments and analyse results in an accepted scientific manner.
Then imagine a fundamental kind of experimenter-subject dependence is discovered that cannot be eliminated and, furthermore, it has occurred in all experiments to date and will occur in all future experiments.
With no way to improve the experiments, there is no reason to reject the results of science to date, nor to abandon practising science in the way it was done.
The consequences of the alternative to experimenter-subject independence would simply be the conditions under which science is done.

Indeed, these conditions include various conservation laws.
In particular, Unnikrishnan has shown that Bell inequalities associated with correlated spin particles are not consistent with the conservation of spin angular momentum, and that satisfying this conservation law requires the Bell inequalities to be violated by an amount given by quantum mechanics \cite{Unnikrishnan2002,Unnikrishnan2005,Unnikrishnan2005a}.
What Bell and Shimony \emph{et al.} \cite{Shimony} refer to as a conspiracy to control ``presumably random choices'' amounts to nothing more than ensuring the conservation of spin angular momentum.

%
%

%
%




\competing{The author declares no competing interests.}

%
%


\end{document}